\documentclass[5p,twocolumn]{elsarticle} %
\usepackage{verbatim}
\usepackage{hyperref}
\usepackage[dvipsnames]{xcolor}
\journal{Microporous and Mesoporous Materials}

\usepackage[]{subfigure}
\usepackage{placeins} 
\usepackage[english]{babel}  
\usepackage[utf8x]{inputenc}
\usepackage{graphicx}
\usepackage{booktabs}
\usepackage{tabularx}
\usepackage{caption}
\usepackage{amsmath}
\usepackage{amssymb}
\usepackage{multirow}

\biboptions{authoryear}

\begin{document}

\begin{frontmatter}

\title{Gas Adsorption and Dynamics in Pillared Graphene Frameworks}

\author[DICAM,TIPFA]{Andrea Pedrielli}

\author[TIPFA,PRAGA]{Simone Taioli\corref{mycorrespondingauthor}}
\cortext[mycorrespondingauthor]{Second corresponding author}
\ead{taioli@ectstar.eu}

\author[TIPFA]{Giovanni Garberoglio\corref{mycorrespondingauthor1}}
\cortext[mycorrespondingauthor1]{First corresponding author}
\ead{garberoglio@ectstar.eu}

\author[DICAM,LONDON,KET]{Nicola Maria Pugno}

\address[DICAM]{Laboratory of Bio-Inspired and Graphene Nanomechanics, Department of Civil, Environmental and Mechanical Engineering, University of Trento, Via Mesiano 77, 38123 Trento, Italy }
\address[TIPFA]{European Centre for Theoretical Studies in Nuclear Physics and Related Areas (ECT*-FBK) and Trento Institute for Fundamental Physics and Applications (TIFPA-INFN), 38123 Trento, Italy}
\address[PRAGA]{Faculty of Mathematics and Physics, Charles University, 180 00 Prague 8, Czech Republic}
\address[LONDON]{School of Engineering and Materials Science, Queen Mary University of London, Mile End Road, London E1 4NS, United Kingdom }
\address[KET]{Ket Lab, Edoardo Amaldi Foudation, Italian Space Agency, Via del Politecnico snc, 00133 Rome, Italy}

\begin{abstract}
 Pillared Graphene Frameworks are a novel class of microporous materials made by graphene sheets separated by organic spacers. One of their main features is that the pillar type and density can be chosen to tune the material properties. In this work, we present a computer simulation study of adsorption and dynamics of H\textsubscript{2}, CH\textsubscript{4}, CO\textsubscript{2}, N\textsubscript{2} and O\textsubscript{2} and binary mixtures thereof, in Pillared Graphene Frameworks with nitrogen-containing organic spacers. In general, we find that pillar density plays the most important role in determining gas adsorption.
 In the low-pressure regime ($\lesssim 10$~bar) the amount of gas adsorbed is an increasing function of pillar density. At higher pressures the opposite trend is observed.
 Diffusion coefficients were computed for representative structures taking into account the framework flexibility that is essential in assessing the dynamical properties of the adsorbed gases. Good performance for the gas separation in CH\textsubscript{4}/H\textsubscript{2}, CO\textsubscript{2}/H\textsubscript{2} and CO\textsubscript{2}/N\textsubscript{2} mixtures was found with values comparable to those of metal-organic frameworks and zeolites. 
\end{abstract}

\begin{keyword}
Nanomaterials\sep  Molecular Dynamics\sep   Gas adsorption\sep  Porosity
\end{keyword}

\end{frontmatter}


\section{Introduction}
In order to exploit graphene for gas adsorption and mechanical applications, many different kinds of 3D carbon-based structures were proposed in the past years, such as carbon nanotube networks \citep{Ding2007}, carbon nanoscrolls \citep{Mpourmpakis2007,Coluci2007} and graphene foams \citep{Alonso2012, Pedrielli2017}. 
At the same time, a growing interest was shown to materials in which graphene is enhanced by chemical functionalization or the addition of external components such as organic molecules~\citep{Tang2013}. In this last category, Pillared Graphene Frameworks (PGF) are a novel class of materials, composed by stacked graphene layers separated by organic moieties. 

\begin{figure}[t]
\centering
\includegraphics[width=%
0.4\textwidth]{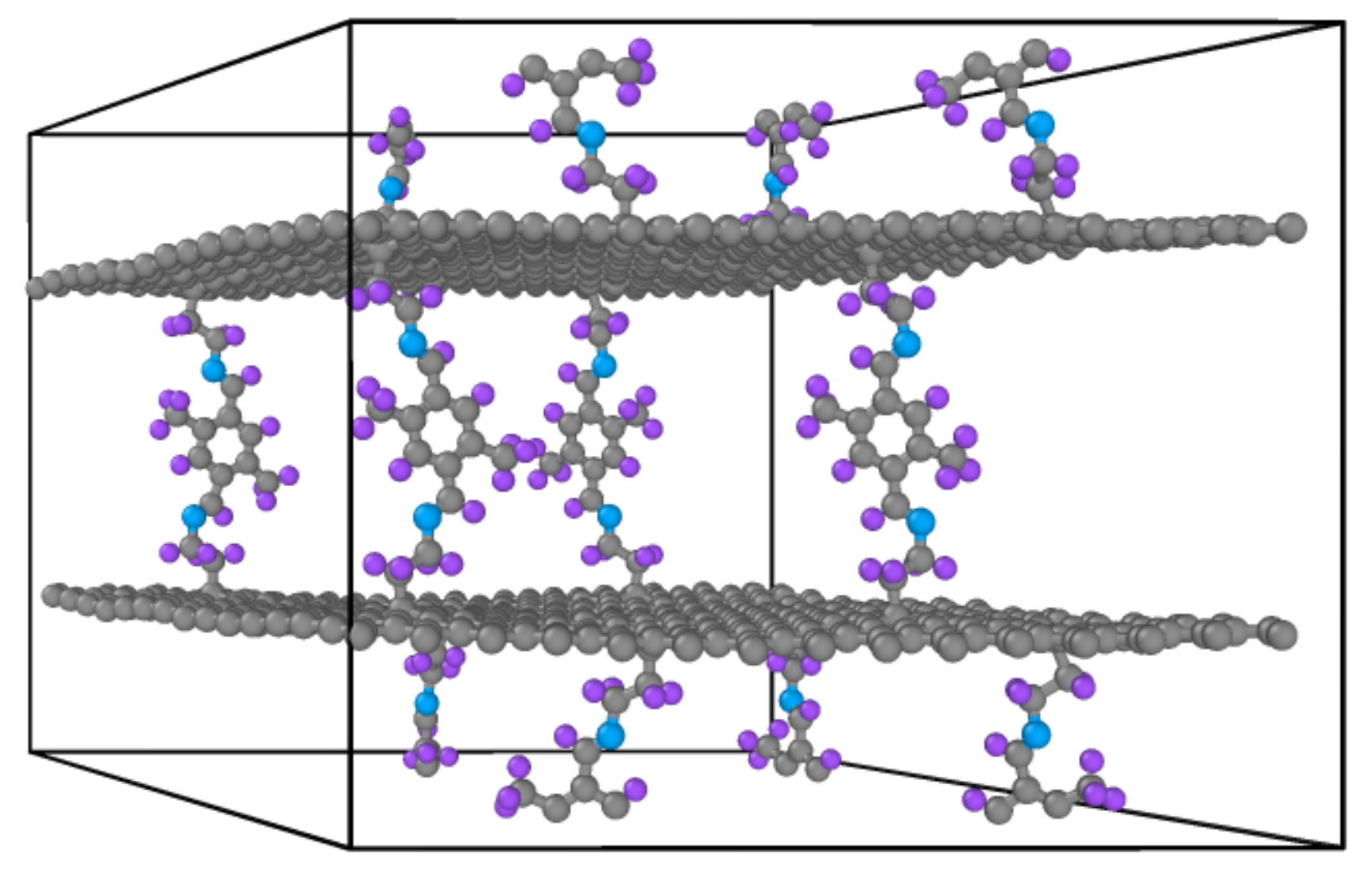}
\caption{View of a Pillared Graphene Framework. The pillars are constituted by organic molecules covalently bonded to graphene layers. Carbon atoms are rendered in grey, hydrogen in violet and nitrogen in blue.}
\label{fig:Framework}
\end{figure}

\begin{figure*}[t]
\centering
\subfigure[Pillar type 1]{
\includegraphics[width=0.20\textwidth]{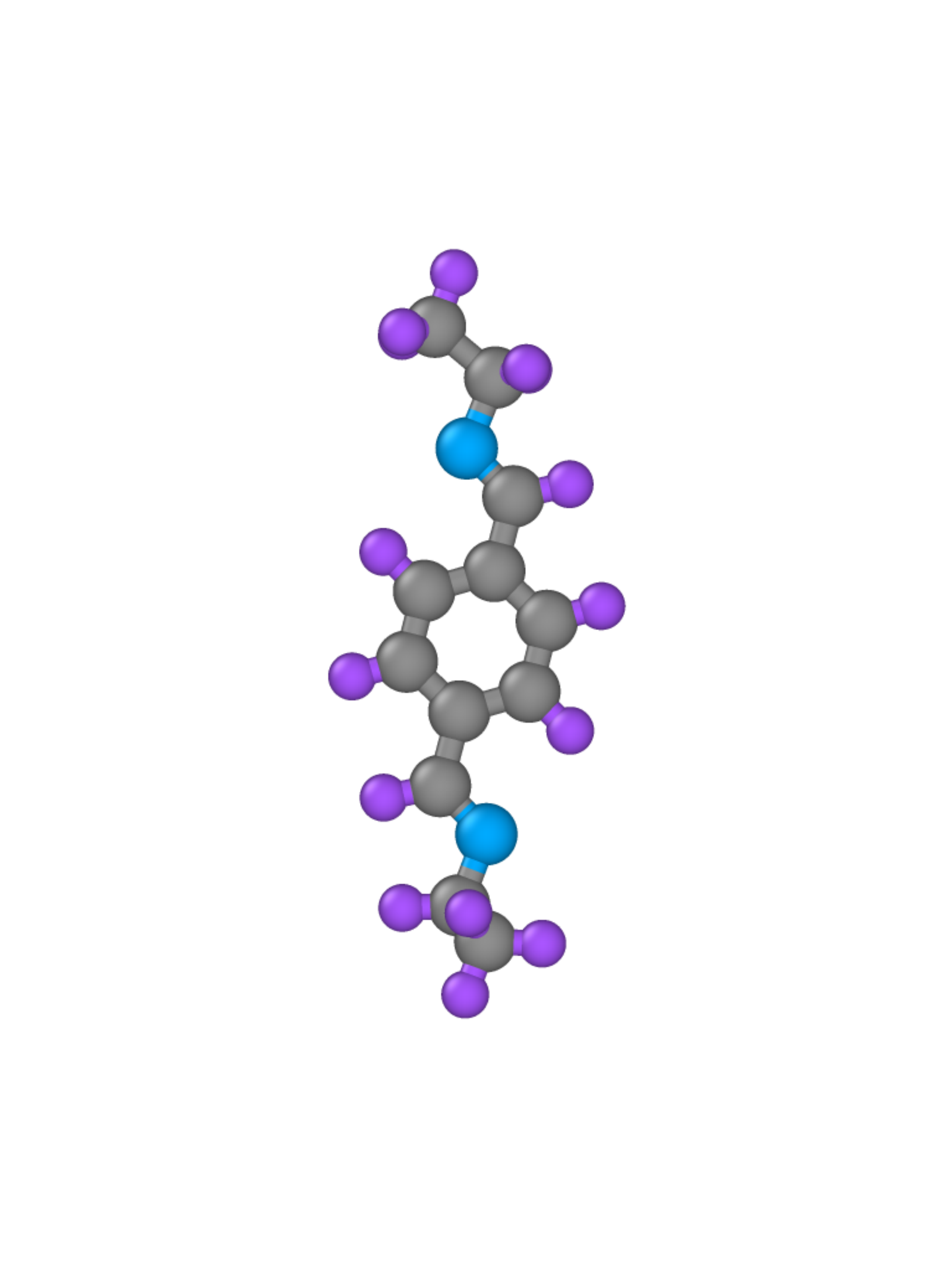}}
\subfigure[Pillar type 2]{
\includegraphics[width=0.20\textwidth]{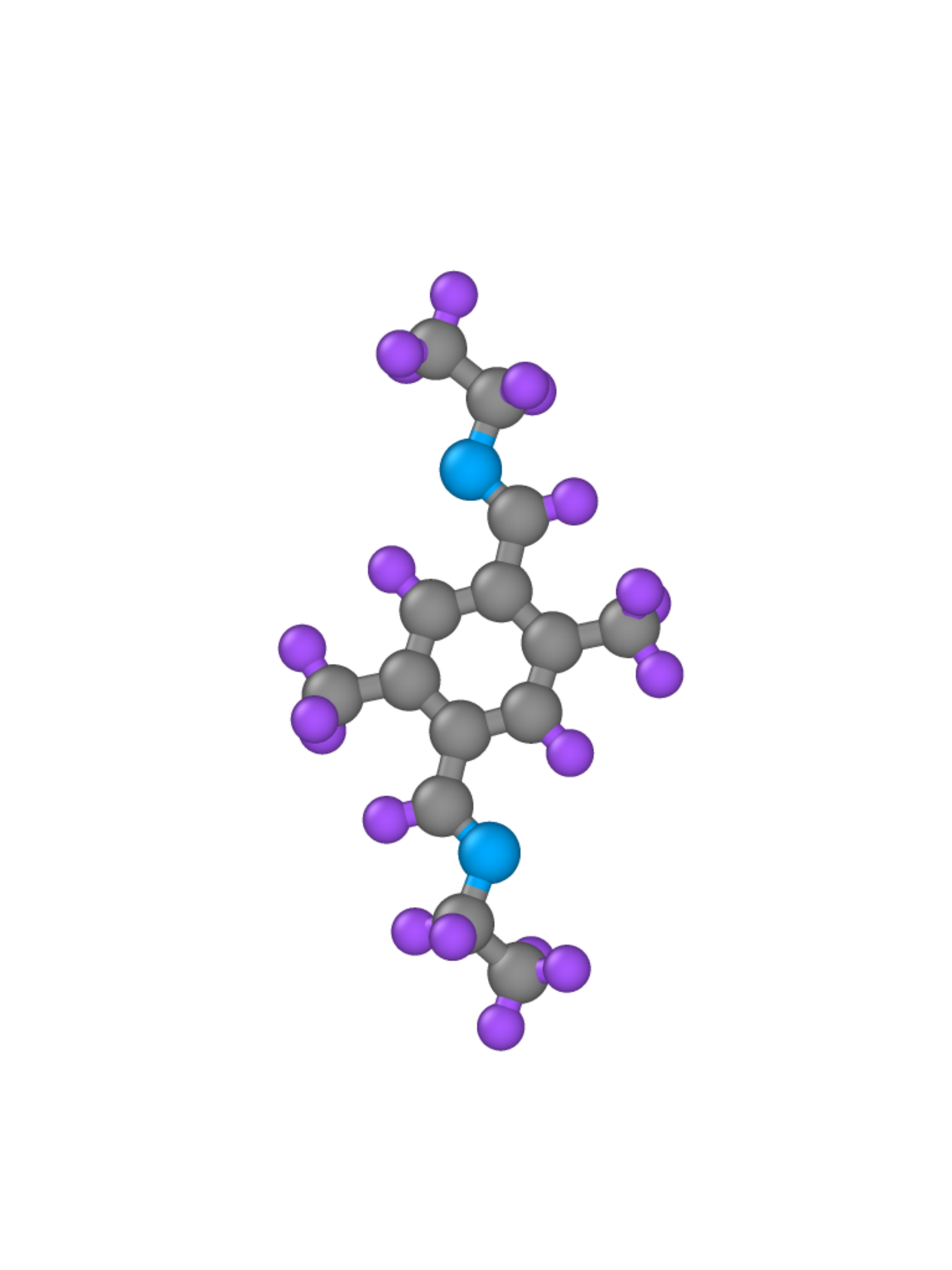}}
\subfigure[Pillar type 3]{
\includegraphics[width=0.20\textwidth]{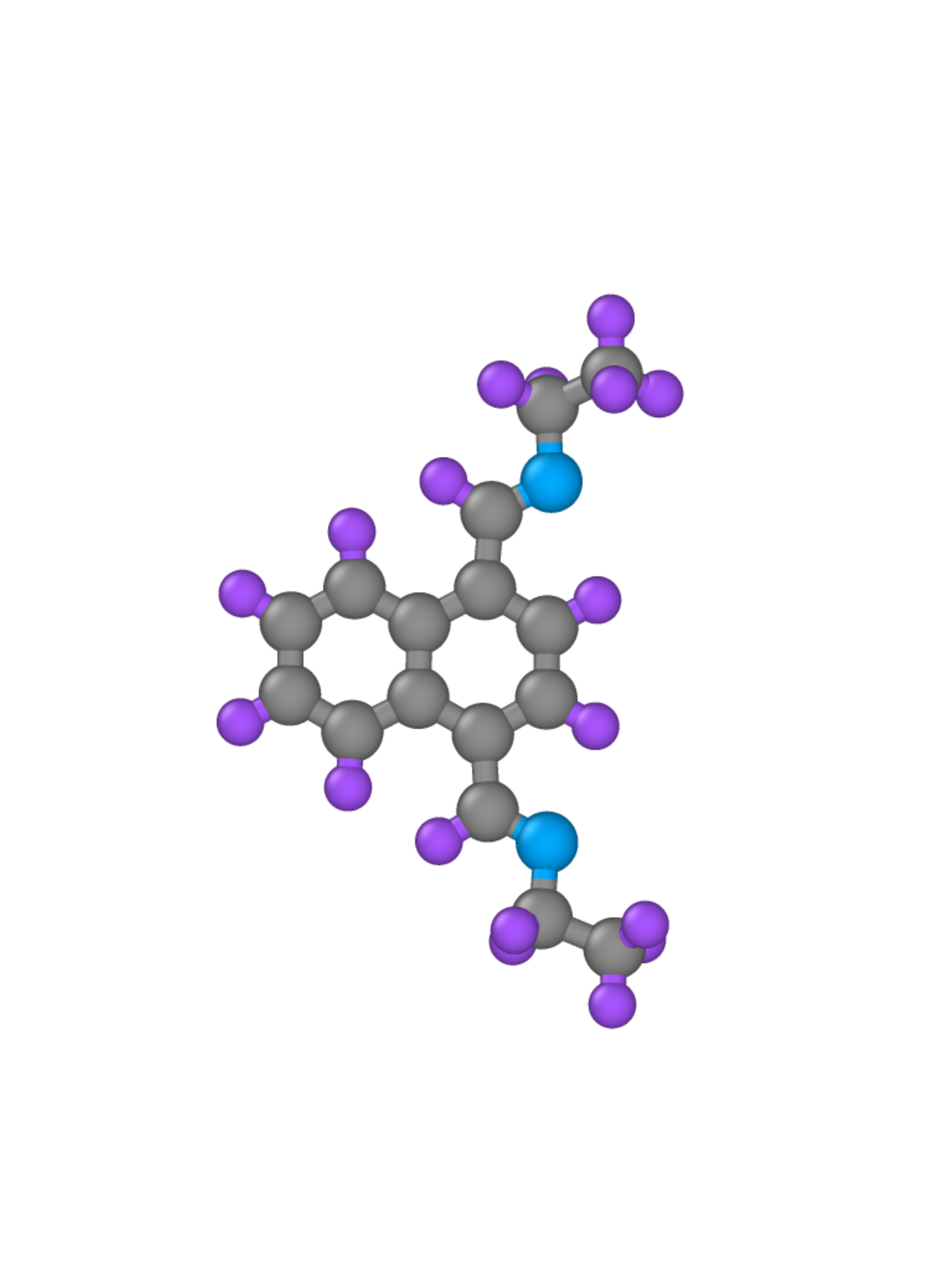}}
\subfigure[Pillar type 4]{
\includegraphics[width=0.20\textwidth]{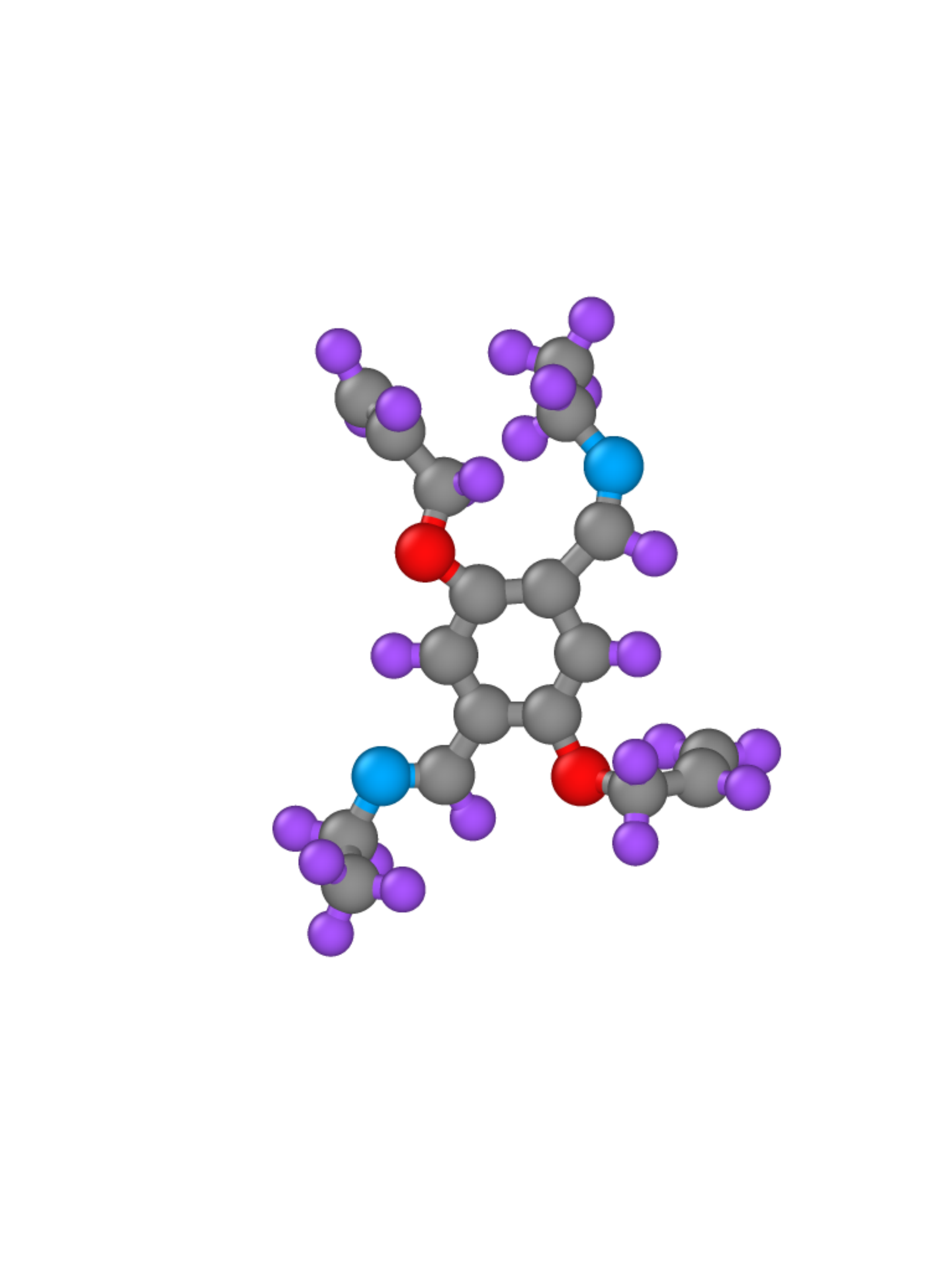}}
\caption{The four nitrogen-containing organic pillars considered in this work. Carbon atoms are rendered in grey, hydrogen in violet, nitrogen in blue and oxygen in red.}
\label{fig:Pillars}
\end{figure*}

Analogously to Pillared Graphene-Oxide Frameworks (PGOF) \citep{Srinivas2011, Kumar2014}, the properties of PGFs can be varied by changing the kind and the density of organic spacers hence obtaining a fine tuning of gas absorption and gas separation performances. Similarly to other materials for gas adsorption such as Metal Organic Frameworks (MOFs), \citep{Duren2004, Babarao2007, Duren2009, Colon2014} Zeolitic Imidazolate Frameworks (ZIFs), \citep{Battisti2011, Zhang2013} and PGOFs \citep{Burress2010,  Garberoglio2015}, the gas adsorption and separation performances of PGFs can be fruitfully studied by means of computer simulations \citep{Wang2014}.

However, gas adsorption and separation in PGFs are still largely unexplored despite their ease of fabrication with respect to other metallic frameworks. In this work, we investigate these properties for a class of structures in which the spacers are nitrogen-containing organic molecules using Grand Canonical Monte Carlo (GCMC) and Molecular Dynamics (MD) simulations. The principal goal of this work is to investigate the role of pillar type and pillar density in determining the performance of PGFs for gas adsorption and gas separation.
In particular, we will investigate whether the quantity of gas adsorbed can be optimized by varying the density of pillars. In fact, one could expect adsorption increase with the number of pillars at small pillar density (providing more adsorption sites), whereas adsorption high pillar density could be prevented by progressive lack of available volume. Consequently, there might be a specific pillar density optimizing gas uptake.

 Furthermore, the influence of pillar density and type on gas separation performances will be assessed. The gas separation performance for a gas mixture depends in general from two main factors: first the gas adsorption of a gas with respect to the other, namely the gas  selectivity, second the difference in the diffusion coefficient of the two species. 
To estimate the gas separation performances of PGFs we will compute the gas selectivity for different mixtures as well as the diffusion coefficients for single component gases.

In computing the diffusion coefficients and assessing the dynamical properties of the adsorbed gases the flexibility of the adsorbent can strongly influence the simulation results, as shown for other materials \citep{Battisti2011, Zhang2014}. Due to the high mobility of the structure considered in this work, we took into account structural flexibility in all dynamical simulations.

\section{Computer model}
Pillared Graphene Frameworks are composed by stacked graphene layers separated by organic spacers. Here we investigate a narrow class of these structures with four types of nitrogen-containing organic spacers. For each type of organic spacer we generated several computational samples with various pillar density, between  $0.09$ and $1.71$ pillars~nm${}^{-2}$. We report in Fig. \ref{fig:Framework} a sketch of an entire sample. In Fig. \ref{fig:Pillars}, the four types of organic pillars considered in this work are shown.

\begin{table}[htbp]
\centering
\small
\begin{tabular}{lccc}
 \\
\toprule 
Pillar   &  Pillar Density &   Free Volume  & Mass Density \\
Type     &    (nm${}^{-2}$) &  (\%) &   (g {cm}$^{-3}$)\\
\midrule 
    &  0.09  &  77.1  &  0.443  \\ 
1   &  0.68  &  67.9  &  0.555  \\ 
    &  1.37  &  57.1  &  0.687  \\ 
\midrule 
    &  0.09  &  76.5  &  0.450  \\ 
2   &  0.68  &  65.8  &  0.569  \\ 
    &  1.37  &  52.7  &  0.721  \\
\midrule 
    &  0.09  &  74.5  &  0.490  \\ 
3   &  0.68  &  65.5  &  0.585  \\ 
    &  1.37  &  51.0  &  0.755  \\
\midrule 
    &  0.09  &  71.0  &  0.560  \\ 
4   &  0.68  &  55.5  &  0.731  \\ 
    &  1.37  &  38.9  &  0.942  \\
\bottomrule
\end{tabular}
\caption{Free volume and mass density for the samples with pillar type $1$ to $4$ and with  representative pillar density. The free volume is defined in Eq. (\ref{eq:FreeVolume}).}
\label{tab:SamplesCharacteristics}
\end{table}

In generating computational supercells, we prepared a hexagonal unit cell with periodic boundary conditions containing two graphene layers with base vectors $a=b=3.684$~nm intercalated by the organic molecules, in such a way that the pillars were alternated in their anchorage to successive graphene planes (see Fig.~\ref{fig:Framework}). The length of the third base vector $c$, perpendicular to the graphene planes, was set to accommodate the pillars, approximately $3$~nm for all the pillar types.
Free volume and mass density for the samples with pillar type $1$ to $4$ and with  representative pillar density are reported in Tab. \ref{tab:SamplesCharacteristics}.

To conclude the preparation of the samples, we equilibrated them using the LAMMPS program \citep{Plimpton1995} by means of $50$~ps isothermal-isobaric Molecular Dynamics simulations at room conditions, using the ReaxFF potential \citep{vanDuin2001,Chenoweth2008} with parameters suitable for organic molecules and carbon-based materials \citep{Mattsson2010}. 
For each sample, we saved one equilibrated configuration of atomic coordinates to be used in the subsequent studies. Furthermore, we saved the point charges that were self-consistently calculated during the ReaxFF simulation (QEq method \citep{Nakano1997, Rappe1991}), 
and we used these point charges in all the simulations in which Coulomb interaction had to be taken into account.

To investigate gas adsorption and separation in these materials we used the Grand Canonical Monte Carlo Method. For a detailed description of the method we refer the reader to a previous paper \citep{Battisti2011}.
In GCMC, as well as in Molecular Dynamics simulations, it is necessary to chose a model for both the gas-gas and the gas-adsorbent interaction. Here we described the molecules as rigid rotors interacting via Lennard-Jones sites and point charges.
In particular we used the EPM2 potential for CO\textsubscript{2} \citep{Harris1995}, and the parameter validated by Murthy for N\textsubscript{2} \citep{Murthy1980} and by Zhang \citep{Zhang2006} for O\textsubscript{2}.  With regard to CH\textsubscript{4} and H\textsubscript{2} we used a single-site Lennard-Jones potential, with the parameters validated by Buch \citep{Buch1994} and Goodbody \citep{Goodbody1991}, respectively. The pure-fluid phase diagram is well described by these models.

The commonly used DREIDING \citep{Mayo1990} force field, augmented with the ReaxFF framework charges, was used to describe the gas-adsorbent interaction. Another popular choice is the UFF force field \citep{Rappe1992}, which we considered for some cases. Analogously to other studies appeared in the literature, we also found that UFF generally results in higher adsorption quantities than DREIDING \citep{Garberoglio2005, Sumida2012, Getman2012}.
The cutoff of the long range van der Waals and Coulomb gas-adsorbent interactions was set to $1.6$~nm.

Framework flexibility could have strong effects on molecular transport in materials with small window sizes or soft components whereas in rigid structures with large pores has minor effects \citep{Amirjalayer2007,Haldoupis2010,Hertag2011,Pantatosaki2012}. 

For the materials considered in this work the pore size as well as the structural rigidity is dependent on the pillar density so that the mobility of the adsorbent during the gas diffusion simulations has to be taken into account. Hence, we used the bonded part of the UFF force field to describe framework flexibility, keeping the DREIDING parameters to describe long-range dispersive interactions. Recent calculations have shown that UFF is very efficient yet capable to describe a broad range of microporous materials with reasonable accuracy \citep{Garberoglio2012A}. 
Indeed, in some preliminary tests, we found that if the framework is kept rigid the underestimation of the diffusion coefficient can be up to $40$~\% lower, when high pillar density structures are considered.
In both GCMC and MD simulations the Lorentz--Berthelot mixing rules were used to calculate the long range van der Waals interaction between unlike atoms.

\section{Results and Discussion}

\subsection{Pure-fluid isotherms}
Pure fluid isotherms were computed for H\textsubscript{2}, CH\textsubscript{4}, CO\textsubscript{2}, N\textsubscript{2} and O\textsubscript{2} gases.
The van der Waals equation of state was used to relate the chemical potential to the pressure of the reservoir gas using parameters set to reproduce the position of the adsorbate critical point \citep{Hirschfelder1954}. For each external pressure we performed $5 \times 10^5$ equilibration steps  (one step being an insertion, a deletion, or a translation/rotation of an already adsorbed molecule, all performed with equal probabilities), followed by 1 million production steps. 

\begin{figure}[htbp]
\centering
\includegraphics[width=0.5\textwidth]{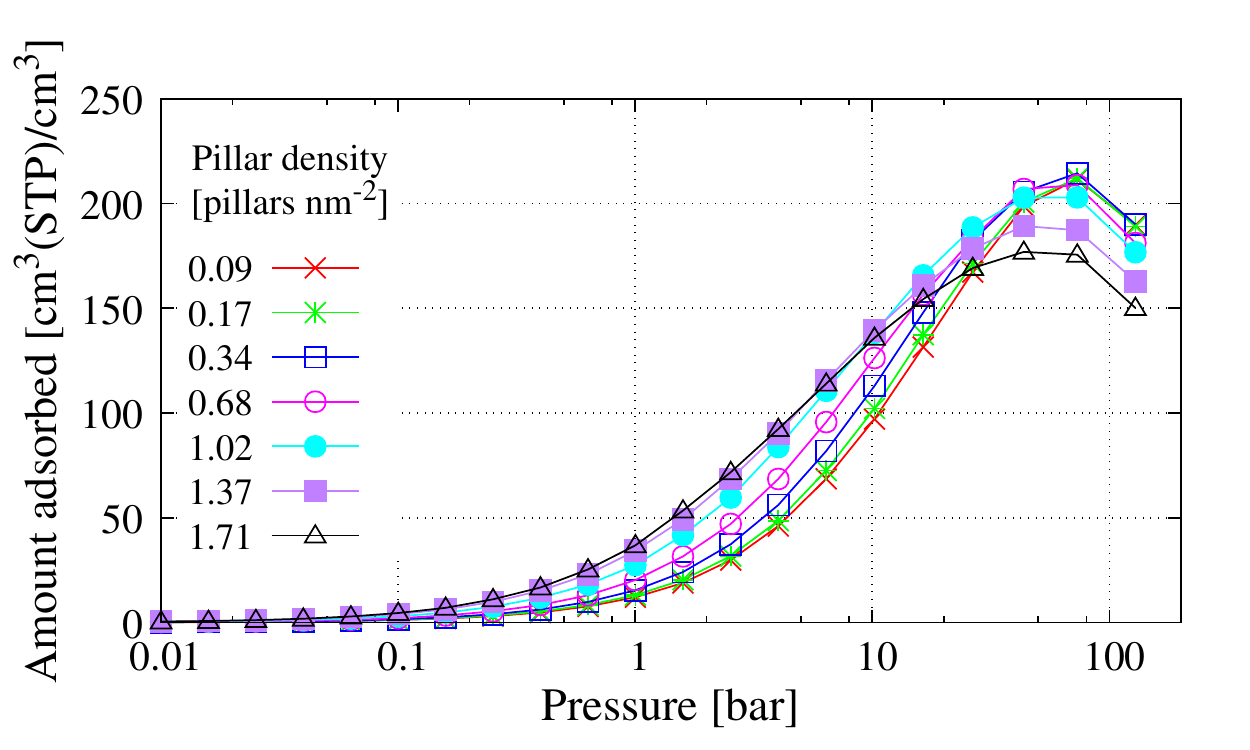}
\caption{Volumetric adsorption isotherms of CH\textsubscript{4} at T= $298$~K for pillar type $1$.}
\label{fig:Iso-Vol-CH4-300}
\end{figure}

\begin{figure}[htbp]
\centering
\includegraphics[width=0.5\textwidth]{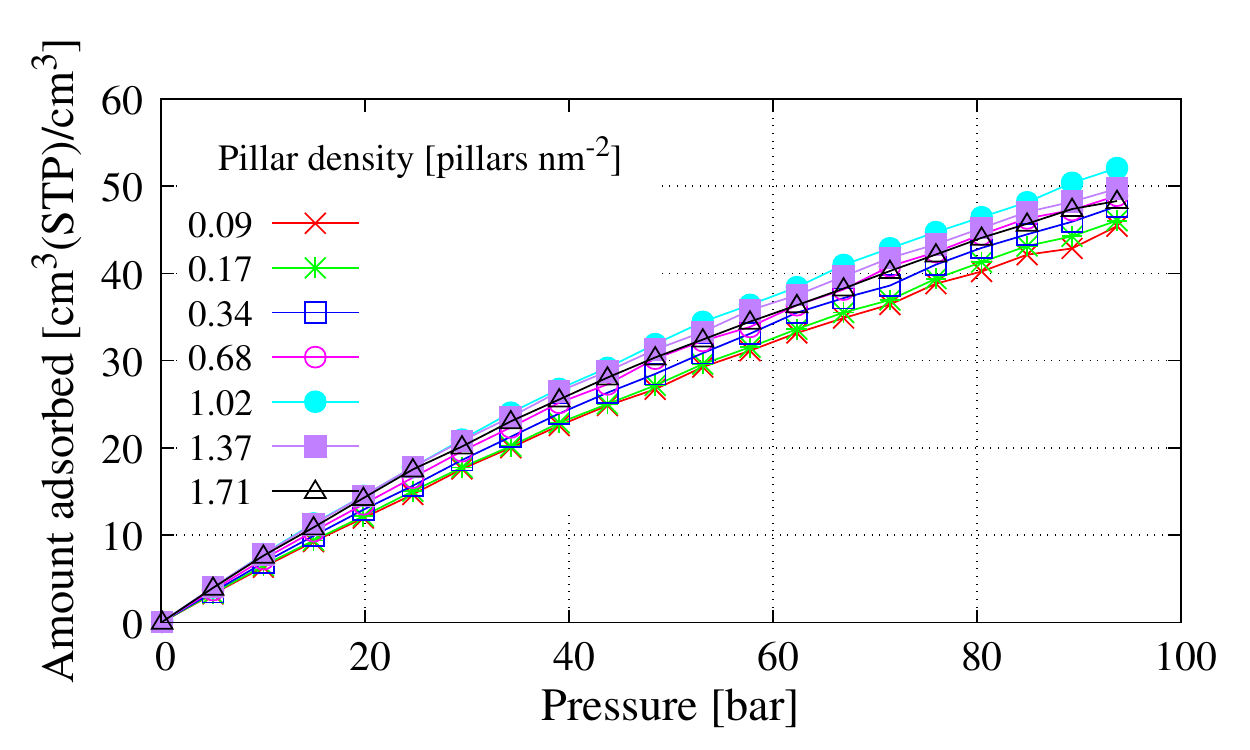}
\caption{Volumetric adsorption isotherms of H\textsubscript{2} at T= $298$~K for pillar type $1$. The best uptake is obtained for an intermediate pillar density of $1.02$ pillars~nm~$^{-2}$.}
\label{fig:Iso-Vol-H2-300}
\end{figure}

\begin{figure}[htbp]
\centering
\includegraphics[width=0.5\textwidth]{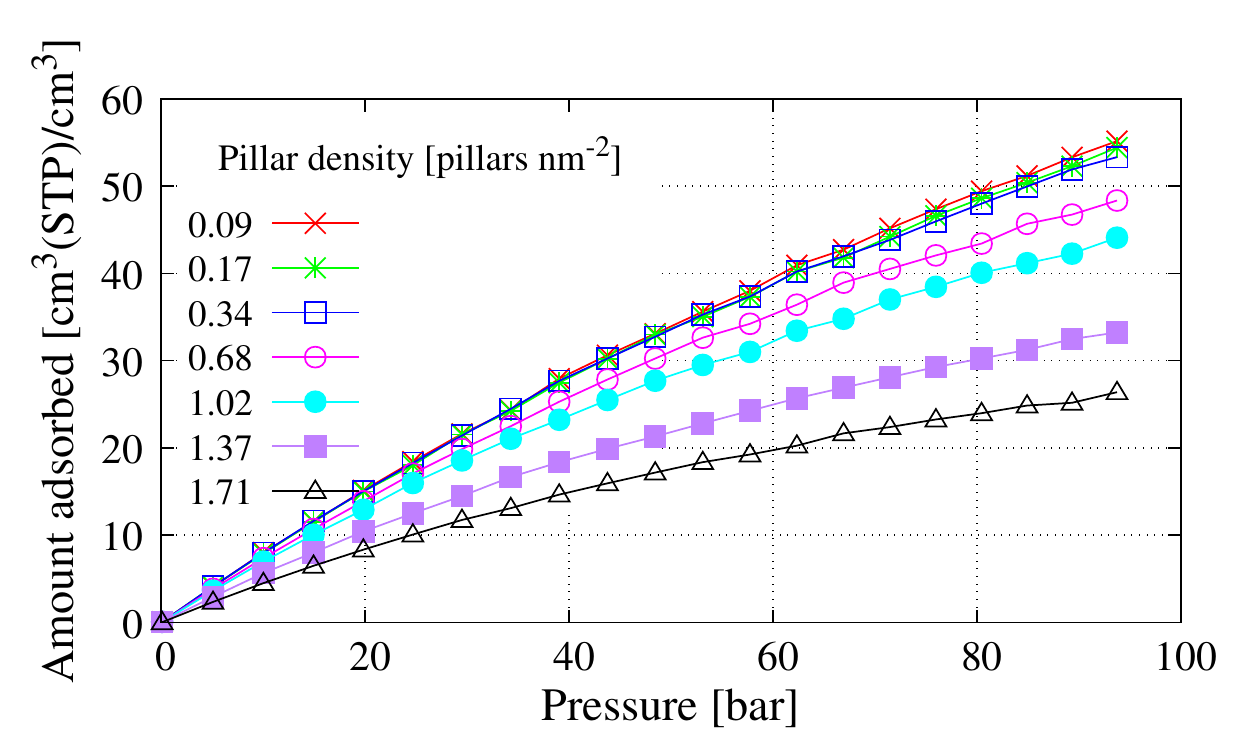}
\caption{Volumetric adsorption isotherms of H\textsubscript{2} at T= $298$~K for pillar type $4$. As the pillar density decreases the adsorption uptake increases.}
\label{fig:Iso-Vol-H2-300-4}
\end{figure}

\begin{figure}[htbp]
\centering
\includegraphics[width=0.5\textwidth]{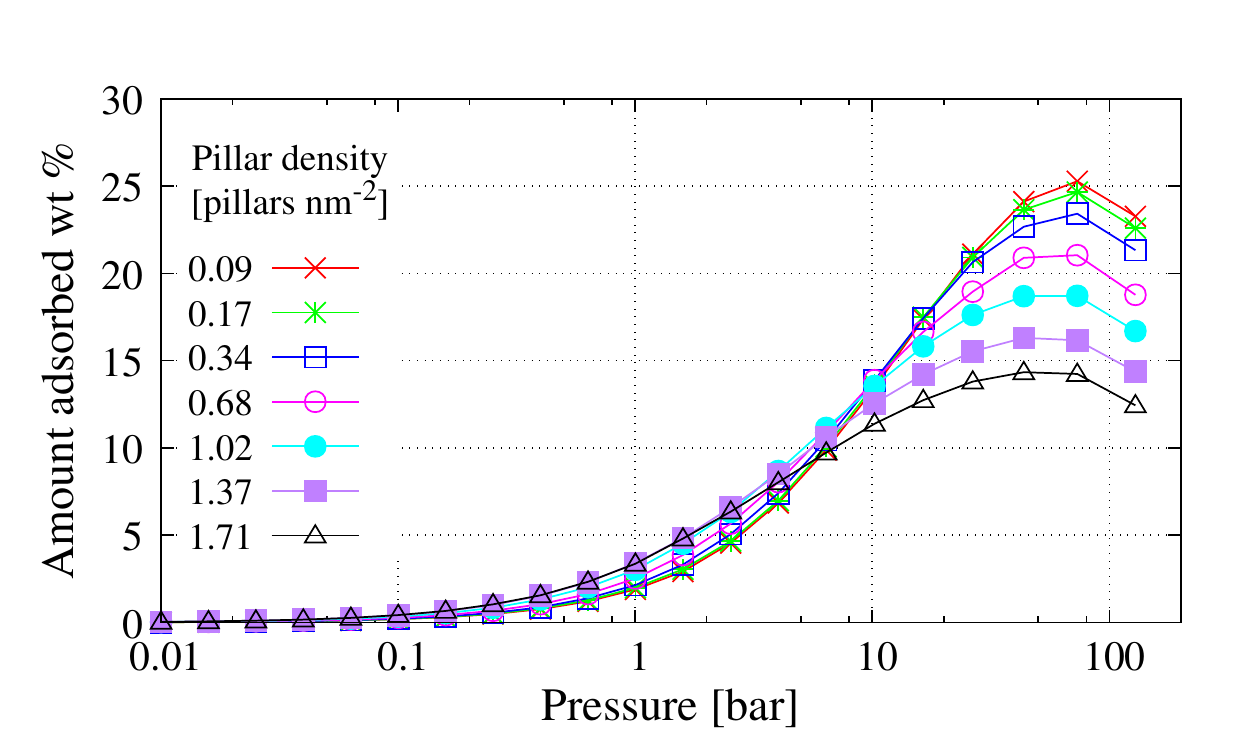}
\caption{Gravimetric adsorption isotherms of CH\textsubscript{4} at T= $298$~K for pillar type $1$. At high pressure, near the saturation limit, we found a clear performance decrease as pillar density increases, the contrary happens at low pressure.}
\label{fig:Iso-Grav-CH4-300}
\end{figure}

\begin{figure}[htbp]
\centering
\includegraphics[width=0.5\textwidth]{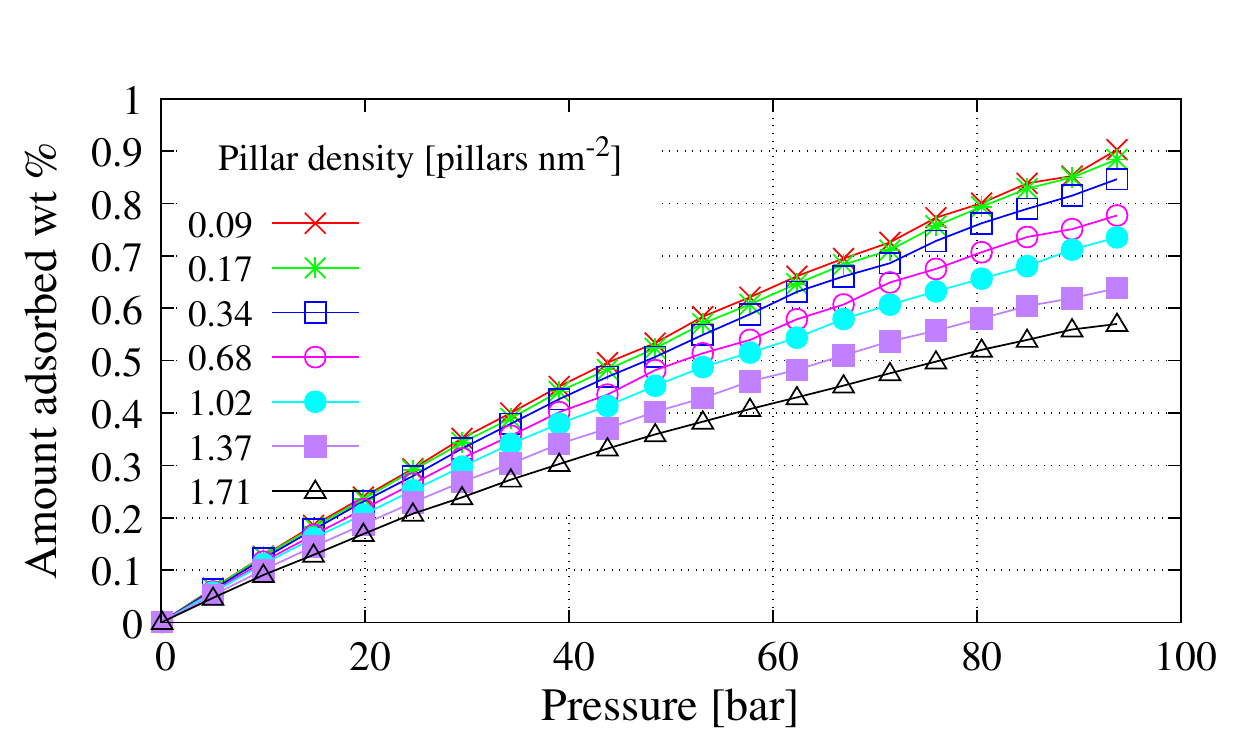}
\caption{Gravimetric adsorption isotherms of H\textsubscript{2} at T= $298$~K for pillar type $1$. As the pillar density decreases the adsorption uptake increases. The saturation is not reached within $100$~bar.}
\label{fig:Iso-Grav-H2-300}
\end{figure}

In particular, we computed the excess amount, $N_\mathrm{ex}$, that can be obtained by estimating the number density $\rho(T,P)$ of the adsorbate at the given thermodynamic condition (calculated using the van der Waals equation of state) and the available free volume for the adsorption  $V_\mathrm{free}$ \citep{Hirschfelder1954}. The free volume is conventionally defined as the volume of the region where the solid-gas interaction between the framework and a helium atom divided by the  Boltzmann constant k${}_\mathrm{B}$ is less than $10^4$~K. 
The excess number of adsorbed molecules is then defined as
\begin{equation}
N_\mathrm{ex} = N - \rho(T,P) ~ V_\mathrm{free},
\label{eq:FreeVolume}
\end{equation}
where $N$ is the total number of gas molecules.


It is in general useful to define two kinds of isotherm curves. 
The first one is the volumetric isotherm which is given by the ratio between the volume occupied by the adsorbed gas at standard pressure and temperature, and the geometric cell volume. This measure of adsorption indicates how much the presence of the adsorbent can concentrate within the adsorbate with respect to room conditions.
The second kind is the gravimetric isotherm and is given by the percent ratio between the weight of the adsorbed gas and the sum of the weights of the framework and the adsorbed gas. This quantity is of practical interest for fuel storage, especially for automotive applications where the weight of the system is of particular concern.

For all the gases (H\textsubscript{2}, CH\textsubscript{4}, CO\textsubscript{2}, N\textsubscript{2}, and O\textsubscript{2}), adsorption isotherms were computed at $298$~K. In the case of H${}_2$ we also considered $T=77$~K.
In what follows, we will focus mainly on isotherms for the pillar type $1$ reporting in the Supplementary Information the results for the other pillar types, because we generally found minor differences as a function of the pillar type. 

Some features of these isotherms are common to almost all the cases investigated in this paper. 
Referring to the volumetric adsorption isotherm of CH\textsubscript{4} at $298$~K reported in Fig.~\ref{fig:Iso-Vol-CH4-300}, one notices that at low pressures (roughly below 10 bar) the quantity of gas adsorbed increases up to two times with increasing pillar density. Indeed, visual inspection of the GCMC configurations shows that in this regime gas is mostly adsorbed close to the framework atoms and a larger number of pillars provides more adsorption sites. This trend was found for all gases execpt H\textsubscript{2} at $298$~K, independently of the pillar type.

Conversely, for larger pressures, the amount of gas adsorbed is a decreasing function of the density of pillars. In this regime, the gas is also adsorbed in the volume between the pillars, but the volume available for adsorption decreases with increasing pillar density due to steric hindrance. Because the maximum volumetric uptake was found for the samples with lower pillar density, the maximum uptake is in general independent of pillar type. In fact, for high pressure, the maximum uptake is essentially limited by the total free volume, that decreases as the pillar density increases.

The volumetric adsorption isotherms of H\textsubscript{2}, reported in Fig. \ref{fig:Iso-Vol-H2-300}, do not follow this general picture. First of all, even at the highest pressure investigated here ($100$~bar) there is no sign of reaching saturation. 

However, despite being in the ``low-pressure regim'', the dependence of the amount adsorbed with respect to the pillar density does not follow the trend observed in the case of the other gases, for one sees that there is an optimal pillar density (around 1~nm${}^{-2}$) that optimizes adsorption, although volumetric uptake is similar (within $20\%$) for all the considered pillar density. The same optimal pillar density was found for PGFs with pillar type $2$ and $3$.
In the case of samples with pillar type 4, reported in Fig. \ref{fig:Iso-Vol-H2-300-4}, this optimal pillar density is not present and we found the uptake being a decreasing function of pillar density. 
This kind of behavior is related to the high pillar volume of the pillar of type $4$, resulting in the lack of free volume also for low pillar density samples.


Gravimetric gas adsorption isotherms at $T=298$~K for the various structures containing pillars of type $1$ and different pillar density are shown in Fig. \ref{fig:Iso-Grav-CH4-300} and \ref{fig:Iso-Grav-H2-300} in the case of CH\textsubscript{4} and H\textsubscript{2}, respectively. 
In the case of CH\textsubscript{4} the isotherms display the same qualitative behavior observed in the volumetric case: adsorption increases with pillar density for low pressures, and decreases at higher ones. However, in this case the normalization with the total mass of the system enhances the difference in adsorption at high pressures, while diminishing it in the low-pressure regime.

For H\textsubscript{2} gravimetric isotherms, reported in Fig.~\ref{fig:Iso-Grav-H2-300} at $298$~K we found, as usual, a linear trend up to $100$~bar, so that saturation is not reached. 
Analogously to methane, when the adsorption per unit mass is considered, higher-density adsorbents are penalized, and in this case the best performance is observed in the lighter structure, independently from the pillar type.

Among the gases considered in this work CH\textsubscript{4}, CO\textsubscript{2} and H\textsubscript{2} are those of major technological interest.
We summarize in Tab.~\ref{tab:UptakeVol} and \ref{tab:UptakeGrav} the maximum values of gravimetric and volumetric uptake found for these gases at $1$, $10$ and $35$ bar, indicating at which pillar type and pillar density corresponds the maximum uptake. 

In the case of CH\textsubscript{4} we found a maximum volumetric uptake at $35$~bar of $195$~cm$^3$(STP)/cm$^3$, with similar performances for different pillar types. 
This value is comparable with what is observed in MOFs, where methane uptake at the same pressure range is $\approx230$~cm$^3$(STP)/cm$^3$ for the best performer \citep{Mason2014}. The performance of the well-known MOF-5 (IRMOF-1) at the same conditions is $\approx 150$~cm$^3$(STP)/cm$^3$.

The amount of CO\textsubscript{2} adsorbed in PGFs is also comparable to what is found in other microporous materials, such as MOFs where gravimetric adsorptions in the range 30--74.2\% are reported at room temperature and pressures up to 50~bar~\citep{Sumida2012}.
The maximum uptake of CO\textsubscript{2} the PGFs examined is reported in Tab.~\ref{tab:UptakeGrav} and can be up to 58.9\% at 35 bar in the case of pillar type 3 at the lowest pillar density.

With regards to H\textsubscript{2} we found a maximum value of $\approx25$~cm$^3$(STP)/cm$^3$ for volumetric uptake at $35$ bar (Tab.~\ref{tab:UptakeVol}) comparable with that of small pore structures such as ZIF-9 and MOF-5 \citep{Garberoglio2005, Battisti2011}. The value for gravimetric maximum uptake of $0.4$\% at $35$ bar, reported in Tab.~\ref{tab:UptakeGrav} is slightly higher than that of MOF-5 and very similar to that of IRMOF-14 \citep{Garberoglio2005}. 

\begin{table*}
\centering
\small
\begin{tabular}{lccc|ccc|ccc}
\toprule 
                   & \multicolumn{3}{c|}{1 bar} & \multicolumn{3}{c|}{10 bar} &\multicolumn{3}{c}{35 bar} \\ 
                   \cmidrule(lr){2-4} \cmidrule(lr){5-7} \cmidrule(lr){8-10}
                   & uptake  &  T & D &  uptake  & T & D & uptake  & T & D \\ 
                   & (cm${}^3$(STP)/cm${}^3$) & & (nm${}^{-2}$) 
                   & (cm${}^3$(STP)/cm${}^3$) & & (nm${}^{-2}$)     
                   & (cm${}^3$(STP)/cm${}^3$) & & (nm${}^{-2}$) \\
\midrule
CH\textsubscript{4}& 43.6 &  2  & 1.71 & 147  &  4   & 0.34 & 195  &  1   & 1.02 \\
CO\textsubscript{2}& 114  &  2  & 1.71 & 341  &  4   & 0.09 & 360  &  3   & 0.09 \\
H\textsubscript{2}& 0.81 &  2  & 0.09 & 8.03 &  2   & 0.09 & 24.6 &  2   & 0.09 \\
\bottomrule
\end{tabular}
\caption{Maximum values of volumetric uptake (cm$^3$(STP)/cm$^3$) found for CH\textsubscript{4}, CO\textsubscript{2} and H\textsubscript{2} at $1$, $10$ and $35$ bar. For each pressure in the last two columns are indicated the pillar type (T) and pillar density (D) producing the maximum uptake.}
\label{tab:UptakeVol}
\end{table*}

\begin{table*}
\centering
\small
\begin{tabular}{lccc|ccc|ccc}
\toprule 
                   & \multicolumn{3}{c|}{1 bar} & \multicolumn{3}{c|}{10 bar} &\multicolumn{3}{c}{35 bar} \\ 
                 \cmidrule(lr){2-4} \cmidrule(lr){5-7} \cmidrule(lr){8-10}
                   & uptake &  T & D &  uptake & T & D & uptake  & T & D \\ 
                   & (wt\%) & & (nm${}^{-2}$) 
                   & (wt\%) & & (nm${}^{-2}$)     
                   & (wt\%) & & (nm${}^{-2}$) \\
\midrule
CH\textsubscript{4}& 3.76 &  2  & 1.37 &  15.2 &  4 & 0.09 & 22.3  &  3   &  0.09 \\
CO\textsubscript{2}& 22.3 &  3  & 1.37 &  53.5 &  3 & 0.09 & 58.9  &  3   &  0.09 \\
H\textsubscript{2}& 0.013 &   3   & 0.09 &  0.13 &  3 & 0.09 & 0.40  &  1   &  0.09 \\
\bottomrule
\end{tabular}
\caption{Maximum values of gravimetric uptake found for CH\textsubscript{4}, CO\textsubscript{2} and H\textsubscript{2} at $1$, $10$ and $35$ bar. For each pressure in the last two columns are indicated the pillar type (T) and pillar density (D) producing the maximum uptake.
}
\label{tab:UptakeGrav}
\end{table*}

\subsection{Comparison between DREIDING and UFF force fields}

As already mentioned, the two force fields that are mostly used to estimate dispersion interactions between adsorbed gases and microporous organic materials are DREIDING and UFF, the latter generally resulting in a higher uptake. In order to compare the results obtained by these two force fields in PGFs, we computed
\begin{equation}
R(P)= \frac{N_\mathrm{ex (UFF)}(P)-N_\mathrm{ex (DREIDING)}(P)}{N_\mathrm{ex (DREIDING)}(P)}, 
\end{equation}
where $N_\mathrm{ex (UFF)}$ and $N_\mathrm{ex (DREIDING)}$ are the excess number of adsorbed molecules at pressure $P$ obtained using UFF force field and DREIDING, respectively. This quantity measures how much adsorption depends on the choice between these two force fields, and is expected to be positive on the basis of the evidence published in literature \citep{Garberoglio2005, Getman2012, Sumida2012}.

\begin{figure}[t]
\centering
\includegraphics[width=0.5\textwidth]{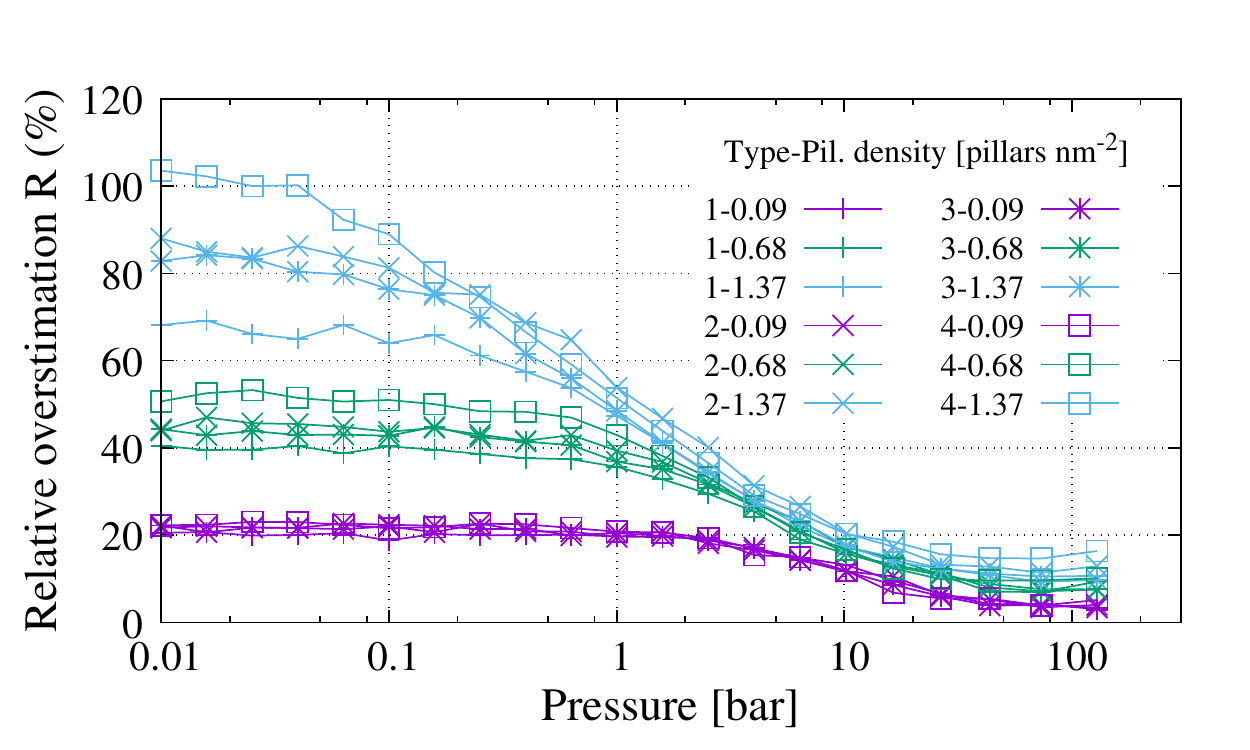}
\caption{Relative overestimation $R$ of CH\textsubscript{4} adsorption at $298$~K using UFF force field in place of DREIDING force field.}
\label{fig:UFF_RelativeCH4}
\end{figure}

The values of $R(P)$ in the case of adsorption of CH\textsubscript{4} at $298$~K temperature for all the pillar types and three different pillar density are reported in Fig.~\ref{fig:UFF_RelativeCH4}, where one can immediately see that also in the case of PGFs UFF predicts a larger amount of gas adsorbed than DREIDING. The various curves present some clear trends. In particular, $R(P)$ is a decreasing function of the external pressure, reaching values less than $20$\% at saturation, and an increasing function of pillar density. This is particularly evident at low pressures ($\lesssim 10$~bar), where UFF predicts up to twice as much adsorbed amount than DREIDING.
In fact, in the low-pressure regime adsorption is mainly determined by the gas-framework interaction, so that the differences between the force fields are emphasized. Conversely, the interaction between gas molecules plays a greater role under saturation conditions (high pressures) and hence the difference due to the two force fields become less important. 
A similar behavior is observed for CO\textsubscript{2} at $298$~K and H\textsubscript{2} at $77$~K. Plots corresponding to Fig.~\ref{fig:UFF_RelativeCH4} can be found in the Supplementary Information.

\begin{figure}[t]
\centering
\includegraphics[width=0.5\textwidth]{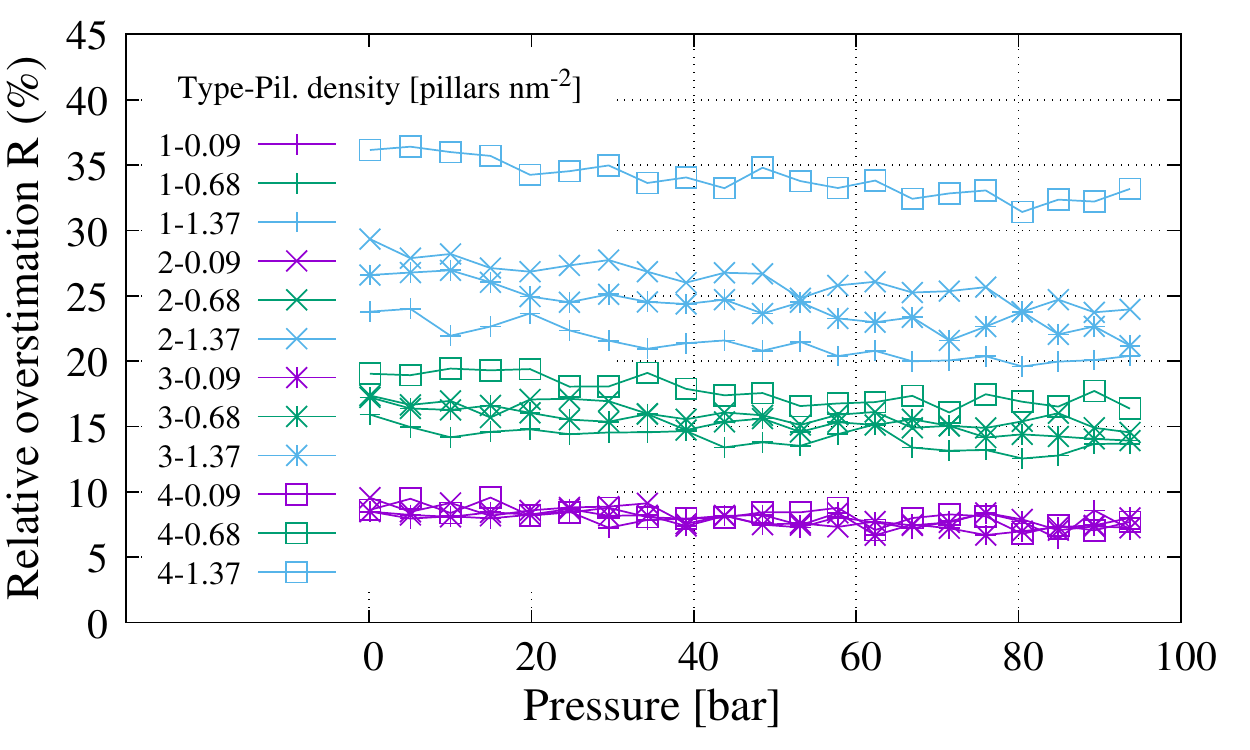}
\caption{Relative overestimation $R$ of H\textsubscript{2} adsorption at $298$~K using UFF force field in place of DREIDING force field.}
\label{fig:UFF_RelativeH2}
\end{figure}

In the case of H\textsubscript{2} at $298$~K instead (Fig.~\ref{fig:UFF_RelativeH2}), $R(P)$ is essentially constant over the whole pressure range, maintaining the dependence on the pillar type and the pillar density found for the previous cases. This can be explained by the fact that the saturation regime is not reached for H\textsubscript{2} at $298$~K, hence the decrease of $R(P)$ at high pressure that is observed in the other gases does not appear in this case.

\subsection{Mixture adsorption and selectivity}

\begin{table}[htbp]
\centering
\small
\begin{tabular}{lcccccccc}
 \\
\toprule 
\multicolumn{1}{l}{Type}  &\multicolumn{3}{c}{ 1} & \multicolumn{3}{c}{ 4} \\ 
\cmidrule(lr){2-4} \cmidrule(lr){5-7}
Density      & \multirow{2}{*}{0.09} & \multirow{2}{*}{0.68} & \multirow{2}{*}{1.37}  & \multirow{2}{*}{0.09} & \multirow{2}{*}{0.68} & \multirow{2}{*}{1.37}   \\ 
 (nm~$^{-2}$)    & &   &  &  & \\
\midrule
CO\textsubscript{2}/H\textsubscript{2}    & 26.2 & 51.4 & 117.6 & 35.8 & 90.2 & 340.0 \\
CH\textsubscript{4}/H\textsubscript{2}    & 9.6  & 16.2 & 31.7  & 12.7 & 24.6 & 66.6  \\
CO\textsubscript{2}/CH\textsubscript{4}   & 2.7  & 3.2  & 3.7   & 2.8  & 3.7  & 5.1   \\
CO\textsubscript{2}/N\textsubscript{2}    & 6.3 & 9.1 & 14.0 & 7.1 & 12.3 & 27.1 \\
N\textsubscript{2}/O\textsubscript{2}     & 1.00 & 1.02 & 1.04  & 1.01 & 1.02 & 1.01  \\
\bottomrule
\end{tabular}
\caption{Zero-pressure adsorption selectivity in the Pillared Graphene Frameworks with pillar types $1$ and $4$ for different pillar density.}
\label{tab:Selectivity}
\end{table}

We investigated the adsorption selectivity of the structures with  pillar types $1$ and $4$ in the case of the following binary mixtures: CH\textsubscript{4}/H\textsubscript{2}, CO\textsubscript{2}/H\textsubscript{2}, CO\textsubscript{2}/CH\textsubscript{4}, N\textsubscript{2}/O\textsubscript{2}, CO\textsubscript{2}/N\textsubscript{2}. We chose to focus on pillar types  $1$ and $4$ that represent the two extrema as pillar complexity and pillar volume: type-1 pillar is linear and not charged, whereas type-4 pillar has protruding charged moieties (see Fig.~\ref{fig:Pillars}).

The selectivity of an adsorbent for a mixture of gases is defined by the ratio 
\begin{equation}
S(b/a)= \frac{x_b/x_a}{y_b/y_a}, 
\end{equation}
where $x_a$, $x_b$ denote the molar fractions of the adsorbed species $a$ and $b$ while $y_a$ and $y_b$ denote the molar fractions of the reservoir bulk mixture. In the low-pressure limit the selectivity is independent of the molar composition of the bulk gas. In this case, it can be computed as the ratio of the single-particle partition function of the two species in the adsorbed phase, divided by the ratio of the free-particle partition function of the same two species \citep{Tan1992, Challa2002, Battisti2011}. 
We denote with $S_0$ the low-pressure limit of the selectivity. 

Values of $S_0$ are report in Tab.~\ref{tab:Selectivity} for the pillar densities of $0.09$, $0.68$ and $1.37$ pillars~nm$^{-2}$, corresponding to the smallest, the intermediate and the higher values investigated in this work. $S_0$ is in general dependent on the considered mixture, the pillar density and the pillar type. The zero-pressure selectivity increases with the pillar density for all mixtures, except N\textsubscript{2}/O\textsubscript{2} for which it is almost constant. The values of the selectivities for these mixtures are generally comparable to the values reported for other microporous materials, such as MOFs~ \citep{Garberoglio2005, Sumida2012}
or ZIFs~\citep{Banerjee2009, Battisti2011, Prakash2013}.

In general, adsorption selectivities in excess of 100 are considered fairly high. In the case of the strucures investigated here, this is observed for the CO\textsubscript{2}/H\textsubscript{2} mixture, especially at high pillar densities where we have $S_0$(CO${}_2$/H${}_2$)$\sim 340$. 
This value is higher than the one found in ZIFs ($\sim 275$~\citep{Battisti2011}) and also in MOFs, where it reaches the value $\sim 100$ in CuBTC and $\sim 12$ in MOF-5~\citep{Yang2006}.

\begin{figure}[t]
\centering
\includegraphics[width=0.5\textwidth]{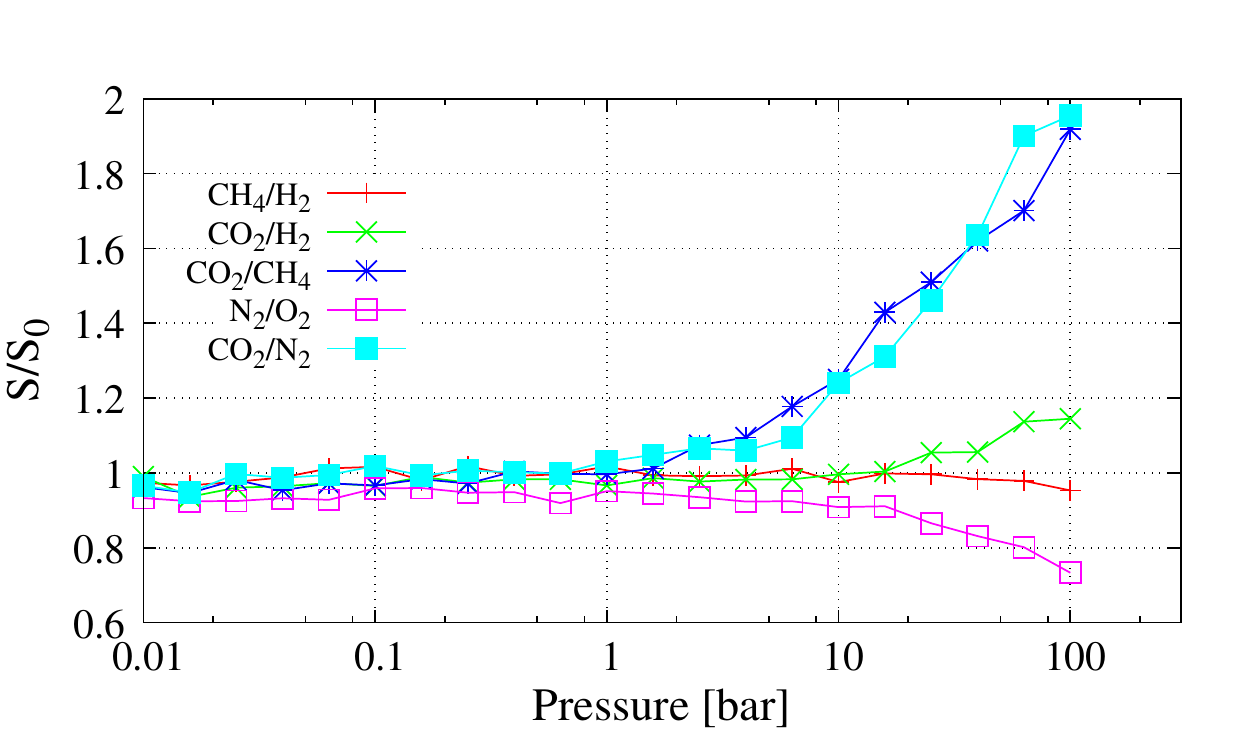}
\caption{Selectivity for gas mixtures at T= $298$~K, normalized with respect to the zero-pressure limit value of selectivity ($S_{0}$), for the sample with pillar type $1$ and pillar density $0.68$ pillars~nm~$^{-2}$. }
\label{fig:Select1}
\end{figure}

\begin{figure}[t]
\centering
\includegraphics[width=0.5\textwidth]{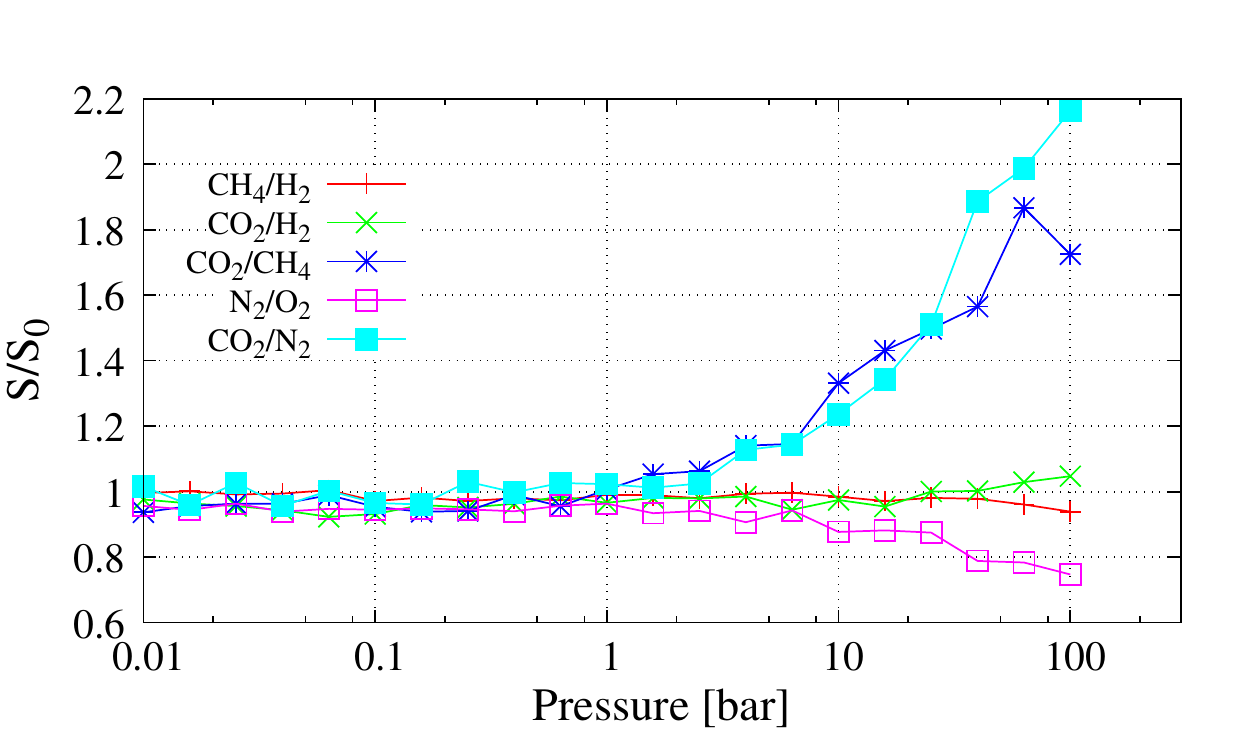}
\caption{Selectivity for gas mixtures at T= $298$~K, normalized with respect to the zero-pressure limit value of selectivity ($S_{0}$), for the sample with pillar type $4$ and pillar density $0.68$ pillars~nm~$^{-2}$.}
\label{fig:Select4}
\end{figure}

As the selectivity is in general a function of the external pressure, its dependence with respect to this parameter was also considered. 
We report in Fig. \ref{fig:Select1} and \ref{fig:Select4} the pressure dependence of $S(b/a)/S_0$ for the samples with pillar type $1$ and $4$, respectively, with a density of $0.68$ pillars~nm$^{-2}$.
As shown in Fig. \ref{fig:Select1} and \ref{fig:Select4} all the mixture selectivities are essentially constant up to $1$~bar keeping their low-pressure value. 
Beyond a few bars we find different trends depending on the mixture: the selectivity can either increase, remain almost constant, or decrease at large pressures with a variation of roughly a factor of two.

The origin of this behavior, which has also been observed in ZIFs \citep{Battisti2011} can be rationalized using energetic and entropic arguments. For molecules of the same type -- e.g. both single Lennard-Jones centers such as CH${}_4$/H${}_2$ or linear rigid rotors such as CO${}_2$/N${}_2$, or N${}_2$/O${}_2$ -- the variation in the selectivity is related to the energetic gain upon adsorption at finite pressure. In general, CO${}_2$ is the molecule whose single-particle energy increases the most when the adsorbed density increases. This in turn enhances the probability of another carbon dioxide molecule being adsorbed with respect to its competing species, resulting in an increasing value of the CO${}_2$ selectivity. This is what happens for the CO${}_2$/N${}_2$ and CO${}_2$/CH${}_4$ mixtures: in both cases the energy gain upon adsorption of a carbon dioxide molecule at the highest pressure is $\sim 200$~K larger than for the adsorption of the other one.
This argument applies also in the case of N${}_2$/O${}_2$, where adsorption of an oxygen molecule results in roughly a $25$~K gain in energy with respect to the adsorption of a nitrogen one. As a consequence, the selectivity decreses at higher pressures.

However, this picture seem to be in contrast with what is observed in the case of the CO${}_2$/H${}_2$ mixture, whose selectivity shows only a modest increase at the highest pressure despite the fact that CO${}_2$ adsorption is favored by $\sim 100$~K gain in energy. In this case one should also take into account the fact that upon adsorption, especially in packed geometries, a carbon dioxide molecule can become rotationally hindered. This loss of entropy balances the gain of energy, resulting in a modest $20$\% gain in selectivity at high pressures. To check this we performed calculations at the lowest and highest pillar densities: in the former case the CO${}_2$/H${}_2$ selectivity increases by up to $50$\%, in the latter it remains constant (within the uncertainties of the calculation).

\subsection{Dynamics of adsorbed molecules and permeation selectivity}

The simulation of gas dynamics was performed using MD simulations within LAMMPS code \citep{Plimpton1995}. The framework topologies to be used in LAMMPS code were generated according to the bonded part of UFF using a modified version of OBGMX~\citep{Garberoglio2012}. We started from representative configuration of pure gas adsorption at two different pressures for which the adsorption uptake was maximum and half maximum. For H\textsubscript{2} at $298$~K we take as maximum the $100$~bar point.

The isothermal simulations started with a $500$~ps equilibration at T = $298$~K driven by a Nos{\'e}--Hoover thermostat with a time constant $\tau=1$~ps. The $x$, $y$ and $z$ components of the mean-squared displacement were computed and averaged over $10$ consecutive $500$~ps MD trajectories, in which the thermostat coupling time was reduced to $\tau=50$~ps. 

The diffusion coefficients were calculated by means of a weighted least square fit of $100$~ps separated points of the averaged mean-squared displacement curve. Indeed the diffusion coefficient can be computed from the mean-squared displacement curve as
\begin{equation}
D_x = \lim_{t \to +\infty} \frac{1}{2} \frac{d\Delta x^2(t)}{dt},
\end{equation}
with an analogous definition for the $y$ and $z$ directions. Due to the fact that all the samples present no gas diffusion in the direction perpendicular to the graphene planes, the overall diffusion coefficient can be assessed as the average between the $x$ and the $y$ directions,
\begin{equation}
D = \frac{1}{2}(D_x+D_y).
\label{eq:D}
\end{equation}

As a first test, we checked the effect of framework flexibility on the values of the self-diffusion coefficient in Eq. (\ref{eq:D}), considering the case of 
CH\textsubscript{4} and CO\textsubscript{2} moving in PGFs with pillar density $0.09$ and $1.37$ pillars~nm${}^{-2}$, pillar type $1$ and $4$, at maximum and half maximum gravimetric uptake.
We found a relative difference in $D$ between mobile and fixed framework of $5-15$~\% and $30-40$~\% for pillar density $0.09$ and $1.374$ pillars~nm${}^{-2}$, respectively. Given these results, we decided to use a flexible model of the framework in the calculation of self-diffusion.

\begin{table}[htbp]
\centering
\small
\begin{tabular}{llcccccc}
 \\
\toprule 
\multicolumn{2}{l}{Type}  & \multicolumn{3}{c}{ 1} & \multicolumn{3}{c}{ 4} \\ 
\cmidrule(lr){3-5} \cmidrule(lr){6-8}
Density      & & \multirow{2}{*}{0.09} & \multirow{2}{*}{0.68} & \multirow{2}{*}{1.37}  & \multirow{2}{*}{0.09} & \multirow{2}{*}{0.68} & \multirow{2}{*}{1.37}   \\ 
 (nm~$^{-2}$)   & & &   &  &  & \\
\midrule
CH\textsubscript{4}&  H  & 30.3 & 12.9  & 3.32  & 35.9   & 7.84  & 0.617 \\ 
                   &  M  & 14.3 & 6.96  & 2.34  & 10.6   & 4.09  & 0.538 \\ 
CO\textsubscript{2}&  H  & 4.04 & 2.85  & 1.27  & 9.56   & 1.60  & 0.270 \\ 
                   &  M  & 0.941& 0.696 & 0.360 & 0.825  & 1.34  & 0.113 \\ 
H\textsubscript{2} &  H  & 219  & 74.3  & 23.6  & 170    & 38.4  & 4.07  \\
                   &  M  & 129  & 49.5  & 20.0  & 105    & 27.8  & 2.79  \\ 
N\textsubscript{2} &  H  & 23.3 & 10.6  & 4.55  & 15.4   & 5.26  & 0.870 \\
                   &  M  & 6.30 & 3.69  & 1.89  & 4.53   & 2.22  & 0.531 \\ 
O\textsubscript{2} &  H  & 28.5 & 11.7  & 5.48  & 23.1   & 5.88  & 1.13  \\
                   &  M  & 8.49 & 4.27  & 2.67  & 5.97   & 2.81  & 0.58  \\ 
\bottomrule
\end{tabular}
\caption{Diffusion coefficients  (in units of $10^{-8}$ m$^2$ s$^{-1}$) for the Pillared Graphene Frameworks with pillar types $1$ and $4$ for different pillar density at half maximum (H) and maximum (M) gravimetric uptake.}
\label{tab:Diffusion}
\end{table}

The results for diffusion coefficient for pillar types number $1$ and $4$ with pillar density $0.09$, $0.68$ and $1.37$ pillars~nm$^{-2}$ are reported in Tab.~\ref{tab:Diffusion}. The general trend is a decrease of the self-diffusion coefficient with increasing pillar density.

Furthermore, H\textsubscript{2} is the gas with higher diffusion values  followed by CH\textsubscript{4}, N\textsubscript{2} and O\textsubscript{2} with similar values, and finally CO\textsubscript{2} with the lower diffusion coefficients.
This sorting is largely independent of the pillar type or density.

The diffusion coefficients reported in Tab.~\ref{tab:Diffusion} are all higher than $10^{-9}$~m$^2$~s$^{-1}$, the order of magnitude of self-diffusion coefficient in liquid such as H\textsubscript{2}O,
so that none of the considered structures inhibits gas diffusion. However, for pillar density higher than $1.37$ pillars~nm$^{-2}$, the gas diffusion could be hindered. 
Differently from ZIFs and MOFs, in which the structures with small windows connecting the pores, such ase, for example, ZIF-5 and ZIF-9 \citep{Battisti2011}, can easily inhibit the gas diffusion, in PGFs the diffusion is not hindered even at high pillar density because the pore are constituted by the free volume between mobile moieties and there are no definite windows to be crossed.

\begin{table}[htbp]
\centering
\small
\begin{tabular}{lcccccc}
\toprule 
\multicolumn{1}{l}{Type}   & \multicolumn{3}{c}{ 1} & \multicolumn{3}{c}{ 4} \\
\cmidrule(lr){2-4} \cmidrule(lr){5-7}
Density       & \multirow{2}{*}{0.09} & \multirow{2}{*}{0.68} & \multirow{2}{*}{1.37}  & \multirow{2}{*}{0.09} & \multirow{2}{*}{0.68} & \multirow{2}{*}{1.37}   \\ 
 (nm$^{-2}$)    & &   &  &  & \\
\midrule
CO\textsubscript{2}/H\textsubscript{2}   & 0.48  & 1.97  & 6.30  &  2.01  &  3.76  &  22.6  \\ 
CH\textsubscript{4}/H\textsubscript{2}   & 1.33  & 2.83  & 4.46  &  2.68  &  5.02  &  10.1  \\ 
CO\textsubscript{2}/CH\textsubscript{4}  & 0.36 & 0.70  & 1.41  &  0.75  &  0.75  &  2.23   \\
CO\textsubscript{2}/N\textsubscript{2}   & 1.09 & 2.44  & 3.89  &  4.38  &  3.75  &  8.39   \\
N\textsubscript{2}/O\textsubscript{2}    & 0.81  & 0.92  & 0.87  &  0.67  &  0.91  &  0.78   \\
\bottomrule
\end{tabular}
\caption{Separation performance factor $\Sigma=\Sigma_0\Pi$ for the Pillared Graphene Frameworks with pillar types $1$ and $4$ for different pillar density.}
\label{tab:Separation}
\end{table}

The overall performance of PGFs for gas separation is determined by a tradeoff between high adsorption selectivity (which is enhanced by high pillar densities, see Tab.~\ref{tab:Selectivity}) and molecular transport (which is hindered by high pillar densities, see Tab.~\ref{tab:Diffusion}). A quantity taking into account this two opposite density regimes is the so called 
permeance selectivity $\Sigma$ which is defined as the product 
\begin{equation}
\Sigma=\Sigma_0\Pi,
\end{equation}
where $\Sigma_0$ is the low-pressure selectivity and $\Pi$ is the ratio between the self-diffusion coefficients of the two gases \citep{Krishna2007,Liu2009,Battisti2011}.

The results for separation performance factor for pillar types number $1$ and $4$ with pillar density $0.09$, $0.68$ and $1.37$ pillars~nm$^{-2}$ are reported in Tab. \ref{tab:Separation}. 
To compute the separation performance factor the diffusion coefficients at half maximum of gravimetric uptake were used.

As general trend the separation performance factor for a given mixture increases as the pillar density increases.
We found good performances for the high pillar density samples for CO\textsubscript{2}/H\textsubscript{2} and CH\textsubscript{4}/H\textsubscript{2}  with maximum values of $22.6$ and  $10.1$, respectively. These values are significantly larger than the ones found in the analysis of gas separation in ZIFs\citep{Battisti2011} where values of $3.42$ and $1.42$ where observed.
Inspection of the values of $S_0$ and $\Pi$ show that the origin of the higher performance of PGFs is mainly due to their larger value of $S_0$, since the ratio of the diffusion coefficient leading to $\Pi$ is roughly the same for PGFs and ZIFs.

A value of $\Sigma = 8.39$ was also found for CO\textsubscript{2}/N\textsubscript{2} separation. For this mixture, ZIFs were found to have a maximum value $\Sigma = 10.4$, in the case of ZIF-4 \citep{Battisti2011}. For this particular mixture, PGFs have a slightly less performing separation behavior, despite having a larger value of $S_0$ ($27.1$ versus $8.2$) due to the fact that the self-diffusion coefficient of N\textsubscript{2} is three times higher than that of CO\textsubscript{2}
in PGFs. In the case of ZIF-4, the value of $\Pi$ turns out to be $\sim 1$\citep{Battisti2011}.

\section{Conclusions}

In this paper we presented an extensive analysis of gas adsorption and separation for nitrogen-containing Pillared Graphene Frameworks using computer simulations. In particular, we focused on the influence of the pillar type and the pillar density on the performance for gas storage and separation.
We took into account the quadrupole moment of CO\textsubscript{2},  H\textsubscript{2} and  O\textsubscript{2} molecules. Furthermore, we used the self consistent point charges extracted by ReaxFF simulations to model the Coulomb interactions between the gases and the frameworks. 

Our results show that the density of pillars has a greater influence on adsorption than the pillar type. Under saturation conditions, the increase of pillar density results in a sensible decrease of the amount of gas adsorbed. Despite this shortcoming, the absolute value of the amount adsorbed is comparable to what is observed in organic frameworks (MOFs, ZIFs or COFs), although it falls short to achieving the performance of the best of them.

In the case of adsorption selectivity, we found that one can have a lot of control on the performance by varying both the pillar type and density. The actual range of variability, though, depends on the specific mixture under consideration. In the case of CO\textsubscript{2}/H\textsubscript{2}, the ratio between the maximum and minimum adsorption selectivity at zero-pressure (see Tab.~\ref{tab:Selectivity}) is more than factor of ten. Conversely, the selectivity of the N\textsubscript{2}/O\textsubscript{2} mixture is always close to one, irrespectively on the nature of the pillar considered or its density.
However, selectivity is in general an increasing function of the pillar density.

When dynamical properties are considered, the effect of pillar density is very pronounced. In general we found roughly an inverse proportionality between the pillar density and the self-diffusion coefficient. This finding paves the way to the possibility of tailoring transport properties to a high degree of precision, possibly up to the ballistic regime. However, there might be issues of stability of the Pillared Graphene Structure at very low pillar densities that will have to be addressed.

Finally, when the overall separation performance $\Sigma$ -- which includes both adsorption and diffusion -- is considered, PGFs show quite a good performance when compared with other microporous materials, especially in the case of the CO\textsubscript{2}/H\textsubscript{2} and CH\textsubscript{4}/H\textsubscript{2} mixtures.

\section*{Acknowledgements}

We thank prof. Marco Frasconi for advice on the kind of moieties to be used as pillars.
N.M.P. is supported by the European Research Council PoC 2015
“Silkene” No. 693670, by the European Commission H2020 under the Graphene  Flagship  Core  1  No.  696656  (WP14 “Polymer Nanocomposites”) and under the Fet Proactive “Neurofibres” No.732344.
S.T and G.G. acknowledge funding from previous WP14 “Polymer Nanocomposites” grant.
Access to computing and storage facilities owned by parties and projects contributing to the Czech National Grid Infrastructure MetaCentrum provided under the programme ``Projects of Large Research, Development, and Innovations Infrastructures'' (CESNET LM2015042), is greatly appreciated ({\tt https://www.metacentrum.cz/en/}).

\bibliography{Bibliografia.bib}

\end{document}